\documentclass[aps,prb,reprint,superscriptaddress,twocolumn,a4paper,fleqn]{revtex4-2}

\usepackage{microtype}
\usepackage[utf8]{inputenc}
\usepackage[T1]{fontenc}
\usepackage{amsmath,amssymb}
\usepackage[pdftex]{graphicx}
\usepackage[hyperfigures]{hyperref}
\hypersetup{
    colorlinks=true,
    linkcolor=blue,
    citecolor=red,
    urlcolor=magenta,
}
\usepackage{xurl}
\usepackage{subcaption} 
\usepackage{chemfig}
\usepackage[version=4]{mhchem}

\usepackage{cleveref}
\usepackage{physics}
\usepackage{bm}
\usepackage{parskip}
\usepackage{booktabs}
\usepackage{float}
\restylefloat{figure}

\begin{document}

\title{Ultraviolet plastic scintillators based on naphthalene-doped polystyrene and polyvinyltoluene}  

\author{Nicolás Agustín Molina}

\affiliation{Departamento Física de la Materia Condensada, Centro Atómico Constituyentes (CNEA), Avenida General Paz 1499, San Martín, 1650, Buenos Aires, Argentina}
\affiliation{Instituto de Nanociencia y Nanotecnología (INN), CONICET-CNEA, Nodo Constituyentes, San Martín, Buenos Aires, Argentina}

\author{Victoria Alejandra Gómez Andrade}
\affiliation{Departamento Física de la Materia Condensada, Centro Atómico Constituyentes (CNEA), Avenida General Paz 1499, San Martín, 1650, Buenos Aires, Argentina}
\affiliation{Instituto de Nanociencia y Nanotecnología (INN), CONICET-CNEA, Nodo Constituyentes, San Martín, Buenos Aires, Argentina}

\author{Javier Martín Abbas}
\affiliation{Gerencia Química, CAC, CNEA-CONICET, Av. General Paz 1499, San Martín, 1650, Buenos Aires, Argentina}

\author{Alan Fuster}
\affiliation{Instituto de Tecnologías en Detección y Astropartículas, Argentina}

\author{Luciano Ferreyro}
\affiliation{Instituto de Tecnologías en Detección y Astropartículas, Argentina}

\author{Martín Alurralde}
\affiliation{Departamento de Energía Solar, Centro Atómico Constituyentes (CNEA), Avenida General Paz 1499, San Martín, 1650, Buenos Aires, Argentina}

\author{Martin Mirenda}
\affiliation{Gerencia Química, CAC, CNEA-CONICET, Av. General Paz 1499, San Martín, 1650, Buenos Aires, Argentina}

\author{María Dolores Pérez}
\affiliation{Departamento Física de la Materia Condensada, Centro Atómico Constituyentes (CNEA), Avenida General Paz 1499, San Martín, 1650, Buenos Aires, Argentina}
\affiliation{Instituto de Nanociencia y Nanotecnología (INN), CONICET-CNEA, Nodo Constituyentes, San Martín, Buenos Aires, Argentina}

\author{Paula Giudici}
\affiliation{Departamento Física de la Materia Condensada, Centro Atómico Constituyentes (CNEA), Avenida General Paz 1499, San Martín, 1650, Buenos Aires, Argentina}
\affiliation{Instituto de Nanociencia y Nanotecnología (INN), CONICET-CNEA, Nodo Constituyentes, San Martín, Buenos Aires, Argentina}

\begin{abstract}
This work reports the fabrication and optical characterization of ultraviolet-emitting plastic scintillators based on polystyrene [\ce{-[CH2-CH(C6H5)]_{n}-}] and polyvinyltoluene [\ce{-[CH2-CH(C6H4CH3)]_{n}-}] doped with different concentrations of naphthalene (Naph). Photoluminescence (PL) measurements show that Naph incorporation enhances the emission intensity, inducing a red shift in the emission wavelength of the polystyrene (Ps), while Naph-doped polyvinyltoluene (PVT) exhibits a bimodal emission with peaks around 335 and 350 nm. Time-resolved photoluminescence (TRPL) reveals fast decay components for the undoped matrices and a dominant slow component of approximately 87 ns in the doped samples. Scintillation light yield measurements indicate moderate performance for the undoped polymers and a significant enhancement upon Naph doping. Proton irradiation experiments reveal a reduction in light yield for all samples, with Naph-doped PVT retaining a larger fraction of its initial light yield compared to Ps-based scintillators, indicating improved radiation tolerance. Overall, these results demonstrate the effectiveness of Naph as a UV emission enhancer and highlight Naph-doped PVT as a promising candidate for compact and radiation-resistant scintillation detectors.
\end{abstract}


\maketitle

\section{Introduction}

Conventional radiation detectors based on silicon face sub-stantial limitations in extreme environments. Their high sensitivity to temperature leads to increased leakage current and reduced signal-to-noise ratio, while prolonged exposure to ionising radiation damages the crystal lattice, degrading charge collection efficiency and shortening device lifetime \cite{radsilicio}. Furthermore, the narrow band gap of silicon ($\sim$1.12 eV) renders it sensitive to visible light, limiting its applicability in spectrally selective detection scenarios.

To overcome these limitations, high electron mobility transistors (HEMTs), such as those based on III-V semiconductors, have been proposed as robust alternatives \cite{Baliga}. Group III-nitride semiconductors, including GaN, AlN, and InN, have become key materials in power electronics and optoelectronic devices—including LEDs, lasers, and HEMTs—due to their wide band gap, high breakdown field, large displacement energy, and exceptional thermal stability \cite{strite}\cite{ganreview}. These properties confer remarkable radiation hardness \cite{electron}, making GaN devices particularly attractive for applications in harsh environments, such as high-energy physics, nuclear science, and space instrumentation \cite{sequeira}. Compared to other semiconductors with wide band gaps, such as silicon carbide (SiC), GaN offers higher electron mobility and improved charge carrier transport characteristics \cite{abid}. 

Despite their numerous advantages, or perhaps precisely because of them, HEMT transistors are very poor particle detectors \cite{electron}. However, their high response to UV radiation \cite{ganuv}  can be exploited if they are combined with scintillators that emit in the UV spectral range compatible with the response window of GaN-based devices. 
From the standpoint of detection efficiency, scintillators and GaN-based transistors offer a naturally complementary functionality: direct detection of ionising particles is intrinsically more efficient in volumetric scintillator materials, where the interaction probability increases with detector thickness. In contrast, HEMTs feature a relatively small active volume, but exhibit high quantum efficiency and fast response when operating as ultraviolet photodetectors \cite{ganuv}.
The introduction of a scintillating layer in front of the GaN transistor is also advantageous in economic terms. Although GaN-based devices exhibit significantly enhanced radiation hardness compared to conventional silicon detectors \cite{electron}, prolonged exposure to ionising particles results in a gradual degradation of the electrical performance of the device \cite{sequeira}. Given the complexity and cost of GaN HEMTs, it is therefore advantageous to employ a scintillator as a protective layer, thereby extending its operational lifetime.

Plastic scintillators are widely recognised for their versatility and effectiveness in detecting charged particles, especially in experiments that demand a fast temporal response. Although their light yield (Y) is typically lower (10,000 photons/MeV compared to 100,000 photons/MeV for many inorganic scintillators), they offer several notable advantages, including short decay times, low production costs, and the ability to be fabricated in a variety of shapes and sizes. Despite the extensive literature on plastic scintillators, the development of materials specifically engineered to emit in the UV-B range ($\sim$280–320 nm) has not been reported to date.

In response to this identified need, the present research undertakes the advancement of plastic scintillators capable of absorbing ionizing particles and re-emitting its energy as UV radiation, tailored to the spectral sensitivity of a GaN-based transistor. By integrating these components, we develop a UV-sensitive, radiation-hardened detection system. In this work, we fabricate the plastic scintillators and characterize their optical properties, estimate their scintillation efficiency, and finally study their degradation under irradiation with 10 MeV protons, energies commonly used in radiation damage experiments for devices intended for space applications \cite{electron}.

\section{Materials and methods}

\subsection{Preparation of plastic scintillators}

Scintillator films were fabricated using Ps and PVT matrices with varying dopant concentrations: pure Ps; Ps loaded with 5, 10, 15, and 20\% w/w Naph; pure PVT; and PVT loaded with 5, 10, 15, and 20\% w/w Naph. For all samples, each gram of polymer was dissolved in 5 mL of toluene per gram of polymer, followed by the addition of Naph under vigorous stirring.

\begin{figure}[h!]
    \centering
    \includegraphics[width=0.9\linewidth]{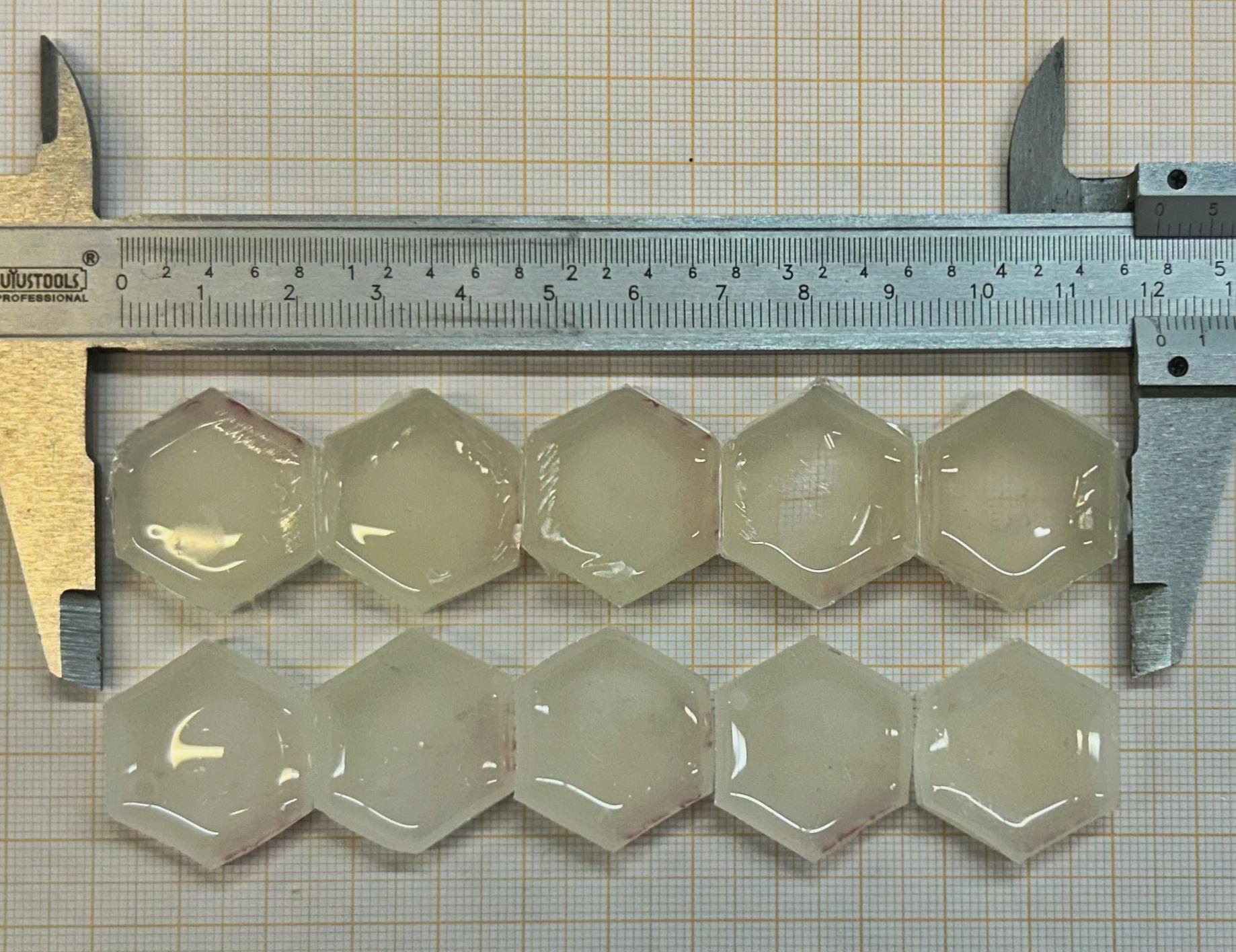}
    \caption{Photograph of Ps-based scintillators doped with Naph (top row) and PVT-based scintillators doped with Naph (bottom row), each set containing samples with 0 wt\%, 5 wt\%, 10 wt\%, 15 wt\%, and 20 wt\% concentrations (from left to right). All samples have a thickness of approximately 2 mm.}
    \label{fig:scints}
\end{figure}

\subsection{Photoluminescence characterization}

PL measurements were performed using a xenon lamp coupled to a monochromator (set at 265 nm) as the excitation source. Emitted light was collected and dispersed by a diffraction-grating spectrometer, with photon energy and intensity distribution recorded by a photomultiplier tube (PMT). Additionally, absorbance spectra of the samples were acquired using a PerkinElmer UV-Vis spectrophotometer to characterize their optical absorption properties in the ultraviolet and visible ranges.

Time-correlated single-photon counting (TCSPC) was employed to measure TRPL \cite{trpl}. The samples were periodically excited using a pulsed laser diode (265 nm), and the emission was collected at the wavelength corresponding to the maximum intensity (\cref{fig:lifetime}). The measured intensity after $P$ excitation pulses corresponds to the sum of the individual decay responses, which can be modeled as a sum of exponential components,

\begin{equation}
I_{TRPL}(t)=\sum_{i=1}^{n}A_{i}e^{-t/\tau_{i}}
\label{eq:ITRPL_exp}
\end{equation}

where $A_{i}$ and $\tau_{i}$ represent the amplitude and the lifetime of the $i$-th component, respectively. The amplitudes correspond to the pre-exponential factor of each decay component and are not intrinsically normalised. The relative weight of each lifetime component is determined after normalisation of the amplitudes. For scintillators, the decay is commonly fitted with bi- or tri-exponential functions, indicating the presence of two or three dominant lifetime components.

\subsection{Scintillation efficiency and light yield}
\begin{figure*}[t!]
    \centering
    \begin{subfigure}[b]{0.48\textwidth}
        \centering
        \includegraphics[width=\linewidth]{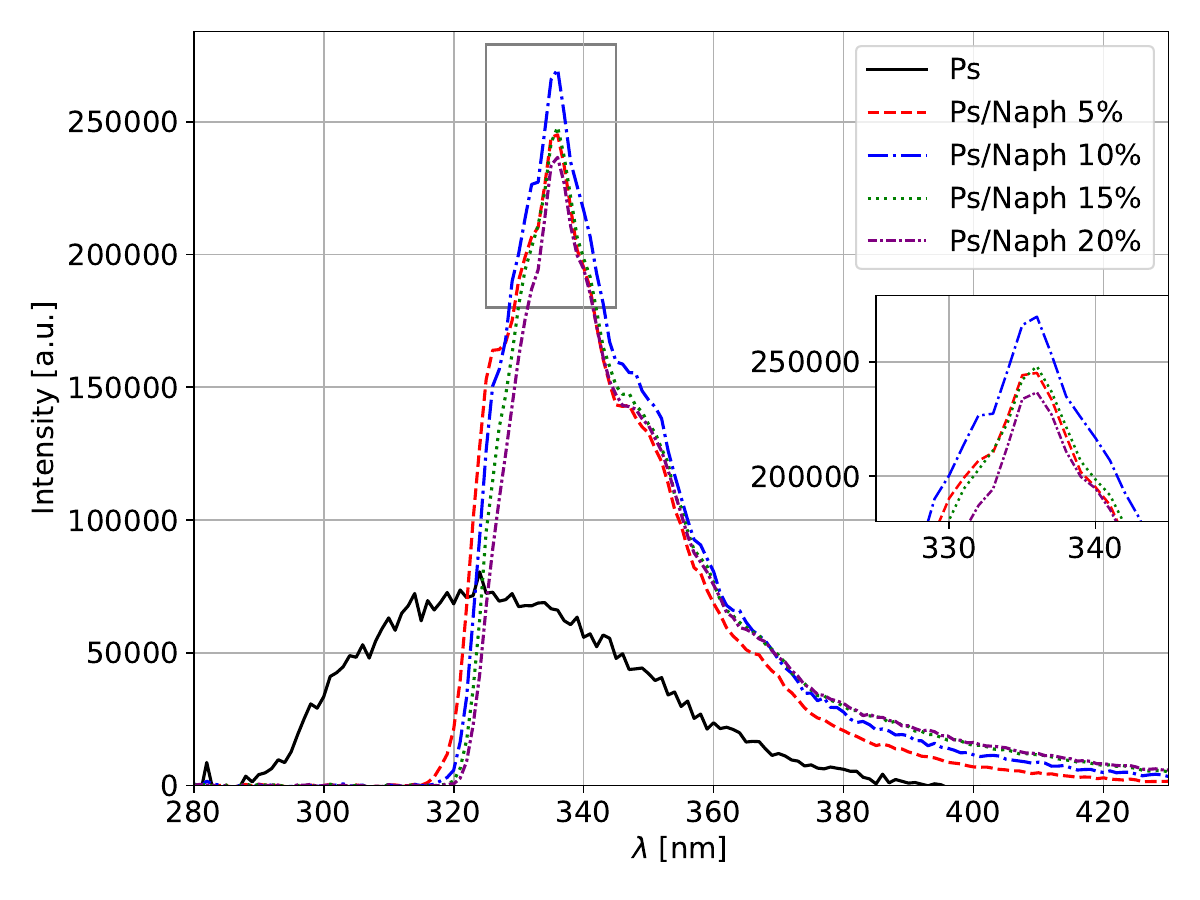}
        \caption{}
        \label{fig:psnaf_sub}
    \end{subfigure}
    \hfill 
    \begin{subfigure}[b]{0.48\textwidth}
        \centering
        \includegraphics[width=\linewidth]{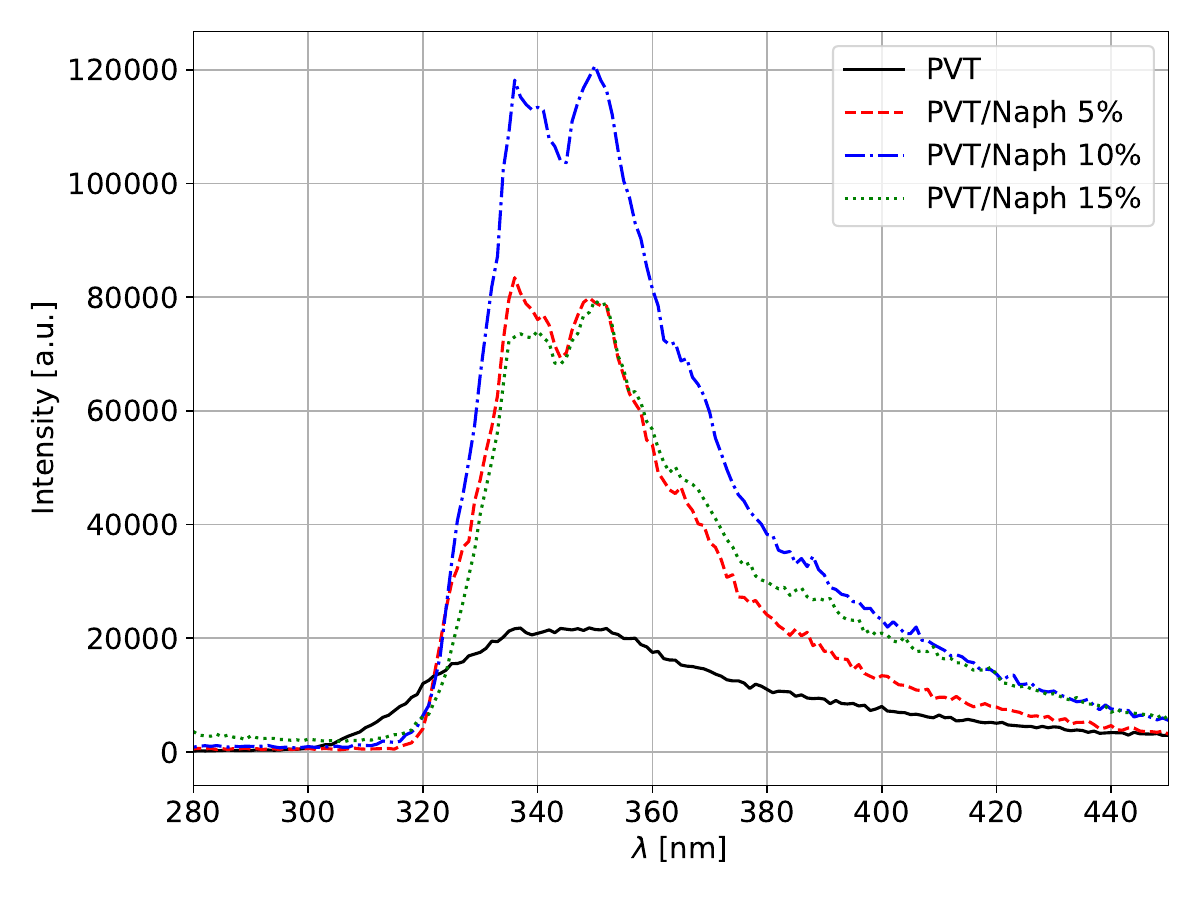}
        \caption{}
        \label{fig:pvt_sub}
    \end{subfigure}
    
    \caption{Photoluminescence spectra at room temperature for (a) Ps and (b) PVT with different loads of Naph. In (a), the inset shows a zoom of the emission maximum region to better distinguish the highest intensity values.}
    \label{fig:combined_loads}
\end{figure*}
The characterization of scintillator light yield was performed by relating the measured charge spectra to the number of detected photons through a calibrated photodetection system. For this purpose, a Hamamatsu PMT \cite{hamamatsu} was calibrated in single photoelectron (SPE) mode to determine the mean charge per detected photon. 

This calibration was performed at 460 nm and subsequently corrected for each scintillator's specific emission spectrum and the PMT’s quantum efficiency. For absolute light yield determination, samples were irradiated with a mixed $\alpha$-source ($^{239}$Pu, $^{241}$Am, and $^{244}$Cm). The total light yield was then calculated by comparing the measured charge from the $\alpha$ particles to the established SPE charge scale. Detailed descriptions of the SPE response modeling, the calibration circuit, and the fitting procedure are provided in the Supplementary Material.

\subsection{Proton irradiation experiments}

Particle irradiation experiments were performed at the TANDAR lineal accelerator of the CAC–CNEA facility \cite{tandar} using a $10$ MeV proton beam.
The proton fluence applied to the samples was monitored in situ using a Faraday cup, allowing precise control of the accumulated charge during irradiation. The target fluence was set to $4.8 \times 10^{11}\ \mathrm{cm^{-2}}$, corresponding to a total absorbed dose of approximately 3 Mrad for each scintillator sample. This irradiation protocol was designed to ensure that all scintillators received comparable radiation doses, enabling a systematic study of radiation-induced degradation effects on their optical and scintillation properties. 

Penetration depth and energy deposition within the scintillator materials of protons were estimated using Stopping and Range of Ions in Matter (SRIM) software. This analysis is essential to ensure that the radiation does not reach or damage any devices placed behind the scintillator, making it critical that the charged particles are fully stopped within the plastic medium. In this context, the use of high-energy protons is particularly relevant, as their interaction with the material may lead to the generation of radiation-induced defects. In our experiments the protons are fully absorved within the scintillators. 

\begin{figure}[h!]
    \centering
    \includegraphics[width=0.5\textwidth]{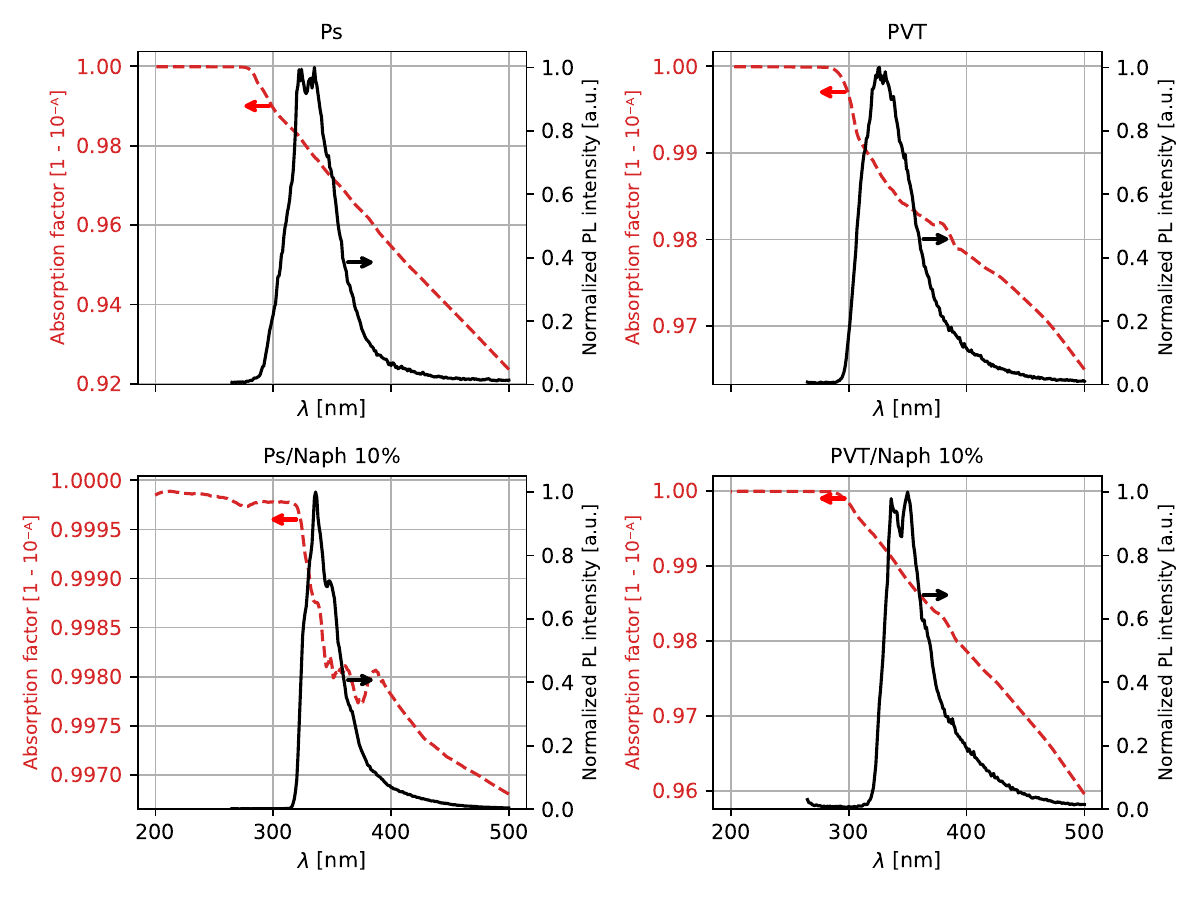}
    \caption{Absortion factor and PL spectra of Ps, Ps doped with Naph, PVT, and PVT doped with Naph. In the red $Y$ labels A is the Absorbance.}
    \label{fig:absypl}
\end{figure}

\section{Results}
\begin{table*}[t!]
\centering
\caption{Characteristic lifetimes ($\tau$) measured via TRPL, emission maxima ($\lambda_\text{max}$) obtained from PL, light yields before irradiation ($Y_0$), and post 3Mrad and 10 MeV proton irradiation ($Y_{\text{irr}}$). Lifetimes are listed in decreasing order of their contributions.}\label{tab:tabla}
\begin{tabular}{@{} l l l l l l l @{}}
\toprule
\textbf{Scintillator} & \textbf{$\tau_1$ (ns)} & \textbf{$\tau_2$ (ns)} & \textbf{$\tau_3$ (ns)} &
\textbf{$\lambda_{\text{max}}$ (nm)} & \textbf{$Y_0$ (ph/MeV)} & \textbf{$Y_{\text{irr}}$ (ph/MeV)} \\
\midrule
Ps & 2.26 (60.0\%) & 11.81 (23.0\%) & 24.76 (17.0\%) & 324 & $1747 \pm 369$ & $1631 \pm 129$ \\
Ps/Naph 10\% & 87.47 (68.80\%) & 2.43 (26.0\%) & 28.75 (5.0\%) & 336 & $5707 \pm 801$ & $3915 \pm 970$ \\
PVT & 8.70 (66.0\%) & 2.47 (19.0\%) & 17.06 (16.0\%) & 326 & $1443 \pm 110$ & $1262 \pm 105$ \\
PVT/Naph 10\% & 88.44 (90.1\%) & 13.41 (9.9\%) & - & 336 & $5287 \pm 780$ & $4481 \pm 589$ \\
\bottomrule
\end{tabular}
\end{table*}

The room-temperature PL spectra of the scintillators with increasing Naph concentrations are presented in \cref{fig:psnaf_sub} (Ps) and in \cref{fig:pvt_sub} (PVT). In the undoped matrices (pure Ps and pure PVT), the emission arises from excimer formation within the polymer. The incorporation of Naph markedly enhances the PL intensity relative to the pure matrices while simultaneously inducing a red shift in the emission peak from 320 nm to 335 nm in the case of Ps. In the Naph-doped samples, the emission is dominated by the Naph monomer. A similar behaviour is observed for PVT; however, the emission spectra of Naph-doped PVT exhibit a double-peak structure, with maxima around 335 nm and 350 nm. This bimodal emission becomes more pronounced as the Naph concentration increases and is accompanied by a strong enhancement of the PL intensity. The emergence of two emission bands highlights a distinct energy transfer and emission mechanism in the PVT-based scintillators.

\begin{figure}[h!]
    \centering
    \includegraphics[width=1\linewidth]{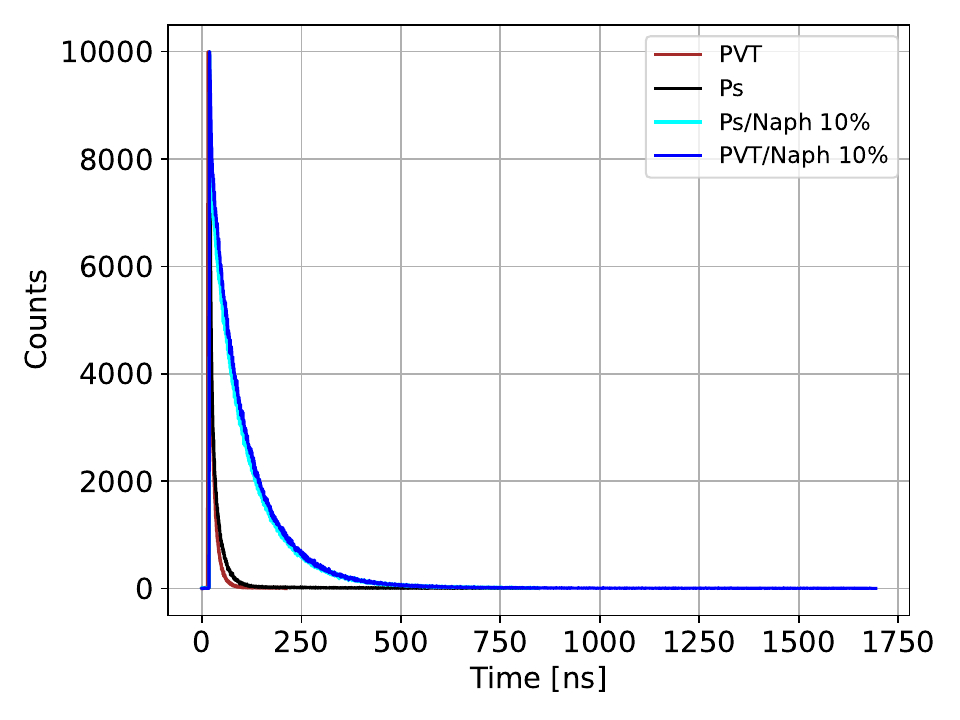}
    \caption{Time-resolved Photoluminescence emission collected at the  wavelength of maximal intensity, exciting with a pulsed laser diode of 265 nm. Significant differences can be observed in the life-time of samples with and without Naphtalene}
    \label{fig:lifetime}
\end{figure}

In \cref{fig:lifetime}, the TRPL curves show that Ps exhibits a significantly faster decay, with a lifetime of about $\sim 2$ ns, whereas the PVT/Naph composite displays a much slower decay, with a characteristic lifetime close to $\sim 87$ ns.

\Cref{fig:absypl} presents the absorbance and PL spectra of the scintillators. The upper panels show the spectra of the undoped Ps and PVT matrices, while the corresponding spectra of the Naph-doped samples are displayed below each one for direct comparison. In all cases, the absorbance spectra remain largely unchanged upon doping, whereas the PL response shows marked differences between doped and undoped scintillators: the addition of Naph to both Ps and PVT matrices results in a pronounced narrowing of the emission spectrum, yielding the characteristic Naph-like band centered at 336 nm and suppressing emission below 310 nm. In contrast, the undoped Ps and PVT samples show broader emission profiles that extend to wavelengths shorter than 300 nm.  

In \cref{tab:tabla}, materials without Naph exhibit dominant decay components in the range of 2 to 9 ns, whereas Ps/Naph and PVT/Naph $10\%$ samples are characterized by a much longer dominant lifetime of approximately 87 ns, attributed to the slower relaxation dynamics of the Naph dopant. While short decay times are advantageous for fast-timing applications, the longer decay constants observed in the doped scintillators are not detrimental for applications such as radiation dose monitoring, where timing resolution is not critical.

Regarding light yield, undoped PVT scintillators exhibit a performance of about $1443\ \mathrm{ph/MeV}$, which is significantly lower than that of commercial scintillators such as EJ-204, whose reported yield is approximately $9000\ \mathrm{ph/MeV}$ \cite{ej204}. Upon Naph doping, the light output of PVT increases substantially, reaching values of up to $5287\ \mathrm{ph/MeV}$, demonstrating a marked improvement with respect to the undoped material. Although this value remains below that of commercial formulations, this difference is consistent with the fact that EJ-204 is an optimized PVT-based scintillator specifically engineered for high efficiency and longer-wavelength emission. Moreover, the light output reported here was obtained from partially opaque samples prepared to prevent saturation of the UV-sensitive phototransistor; this experimental constraint is not intrinsic to the material and can be readily improved through optimized drying procedures.

Light yield experiment was repeat after irradiating the scintillators with 10 MeV protons at 3Mrad. Experiments reveal a degradation of light yield for all samples after exposure, with the extent of the damage depending on both the polymer matrix and the presence of the dopant. In particular, Naph-doped PVT retains a larger fraction of its initial light yield after irradiation compared to the Ps-based counterpart, indicating improved radiation tolerance.

Overall, these results show that Naph-doped PVT combines enhanced UV emission, suitable decay characteristics for the employed detection scheme, and improved resistance to proton-induced degradation, making it a promising candidate for compact scintillation detectors coupled to GaN-based photodetectors.

\section{Conclusions}

Undoped Ps and PVT exhibit moderate light yields, while Naph doping leads to a significant enhancement of the scintillation performance in both matrices. This behavior is consistent with the PL spectra and decay dynamics, indicating that the emission in the doped samples is largely dominated by the Naph component. Although undoped Ps slightly outperforms undoped PVT, both hosts benefit strongly from dopant incorporation, confirming the effectiveness of Naph as a UV emission enhancer in aromatic polymer scintillators.

Proton irradiation experiments reveal that all scintillators undergo a degradation of light yield after exposure. However, the extent of this degradation depends on both the polymer matrix and the presence of the dopant. In particular, Naph-doped PVT retains a larger fraction of its initial light yield after irradiation compared to its Ps-based counterpart, indicating improved radiation tolerance.

Beyond light yield, decay dynamics play a key role in defining suitable applications. Undoped Ps and PVT exhibit short decay times that are favorable for fast-timing detectors, whereas Naph-doped samples are dominated by a slow component. In the present system, this does not represent a limitation, as the response time of the GaN phototransistor used for detection is significantly slower than the scintillator decay.

Overall, the results indicate that Naph-doped PVT provides an attractive combination of enhanced UV emission, acceptable timing characteristics, and improved resistance to proton-induced degradation, making it a promising candidate for compact scintillation detectors coupled to GaN-based photodetectors.

\begin{acknowledgments}
This work was partially funded by the grants PIP 2023-2025 GI (11220220100489CO) and PICT-2021-GRF-TII-
00304. Sincere gratitude is extended to Andrés Di Donato for his continued support and to TANDAR accelerator operators for their patience and exceptional kindness throughout the experimental process.  
\end{acknowledgments}

\bibliography{bib}

@article{ganreview,
    author = {Meneghini, Matteo and De Santi, Carlo and Abid, Idriss and Buffolo, Matteo and Cioni, Marcello and Khadar, Riyaz Abdul and Nela, Luca and Zagni, Nicol\`o and Chini, Alessandro and Medjdoub, Farid and Meneghesso, Gaudenzio and Verzellesi, Giovanni and Zanoni, Enrico and Matioli, Elison},
    title = {GaN-based power devices: Physics, reliability, and perspectives},
    journal = {Journal of Applied Physics},
    volume = {130},
    number = {18},
    pages = {181101},
    year = {2021},
    month = {11},
    issn = {0021-8979},
    doi = {10.1063/5.0061354},
    url = {https://doi.org/10.1063/5.0061354},
    eprint = {https://pubs.aip.org/aip/jap/article-pdf/doi/10.1063/5.0061354/19876866/181101\_1\_5.0061354.pdf},
}

@manual{hamamatsu,
  author       = {{Hamamatsu Photonics K.K.}},
  title        = {NEW COMPACT TYPE PMT SERIES R6350 to R6358},
  type         = {Technical data sheet},
  number       = {R6350\_05},
  year         = {},
  url          = {https://www.alldatasheet.com/datasheet-pdf/pdf/212285/HAMAMATSU/R6350\_05.html},
  note         = {Accedido el 22 de abril de 2026. Serie compacta de 13 mm, tipo side-on, que incluye los modelos R6350 a R6358.},
  organization = {Hamamatsu Photonics K.K.}
}

@article{led,
title = {Multichannel pulsed light source with LED for photomultiplier testing},
journal = {Nuclear Instruments and Methods in Physics Research Section A: Accelerators, Spectrometers, Detectors and Associated Equipment},
volume = {972},
pages = {164058},
year = {2020},
issn = {0168-9002},
doi = {https://doi.org/10.1016/j.nima.2020.164058},
url = {https://www.sciencedirect.com/science/article/pii/S0168900220304927},
author = {Agustín Lucero and Federico Suarez and Carlos Reyes and Alberto Etchegoyen}
}

@misc{tandar,
  author       = {{Comisión Nacional de Energía Atómica (CNEA)}},
  title        = {Acelerador Tandem Argentino (TandAr)},
  howpublished = {Sitio web de Argentina.gob.ar},
  year         = {},
  url          = {https://www.argentina.gob.ar/cnea/cac/tandar},
  note         = {Accedido el 22 de abril de 2026}
}

@Article{trpl,
author ="Péan, Emmanuel V. and Dimitrov, Stoichko and De Castro, Catherine S. and Davies, Matthew L.",
title  ="Interpreting time-resolved photoluminescence of perovskite materials",
journal  ="Phys. Chem. Chem. Phys.",
year  ="2020",
volume  ="22",
issue  ="48",
pages  ="28345-28358",
publisher  ="The Royal Society of Chemistry",
doi  ="10.1039/D0CP04950F",
url  ="http://dx.doi.org/10.1039/D0CP04950F"}

@article{electron,
doi = {10.1149/2.0251602jss},
url = {https://dx.doi.org/10.1149/2.0251602jss},
year = {2015},
month = {nov},
publisher = {The Electrochemical Society},
volume = {5},
number = {2},
pages = {Q35},
author = {Pearton, S. J. and Ren, F. and Patrick, Erin and Law, M. E. and Polyakov, Alexander Y.},
title = {Review—Ionizing Radiation Damage Effects on GaN Devices},
journal = {ECS Journal of Solid State Science and Technology},
}

@article{radsilicio,
    author = "Oblakowska-Mucha, A.",
    collaboration = "RD50",
    title = "{Radiation Damage in Silicon Particle Detectors in High Luminosity Experiments}",
    doi = "10.5506/APhysPolB.48.1707",
    journal = "Acta Phys. Polon. B",
    volume = "48",
    pages = "1707--1714",
    year = "2017"
}

@article{ganuv,
title = {Unveiling the abnormal response behavior of AlGaN-based high electron mobility transistors (HEMTs) under ultraviolet light illumination},
journal = {Materials Science in Semiconductor Processing},
volume = {187},
pages = {109134},
year = {2025},
issn = {1369-8001},
doi = {https://doi.org/10.1016/j.mssp.2024.109134},
url = {https://www.sciencedirect.com/science/article/pii/S1369800124010308},
author = {Mukta Sharma and Chia-Lung Tsai and S.N. Manjunatha and Yu-Li Hsieh and Atanu Das and Kuan-Ying Lee and Sun-Chien Ko and Shiang-Fu Huang and Liann-Be Chang and Meng-Chyi Wu}
}

@article{ej204,
title = {Absolute light yield of the EJ-204 plastic scintillator},
journal = {Nuclear Instruments and Methods in Physics Research Section A: Accelerators, Spectrometers, Detectors and Associated Equipment},
volume = {1054},
pages = {168397},
year = {2023},
issn = {0168-9002},
doi = {https://doi.org/10.1016/j.nima.2023.168397},
url = {https://www.sciencedirect.com/science/article/pii/S016890022300387X},
author = {J.A. Brown and T.A. Laplace and B.L. Goldblum and J.J. Manfredi and T.S. Johnson and F. Moretti and A. Venkatraman}
}

@article{rossipmt,
title = {Methods for precise photoelectron counting with photomultipliers},
journal = {Nuclear Instruments and Methods in Physics Research Section A: Accelerators, Spectrometers, Detectors and Associated Equipment},
volume = {451},
number = {3},
pages = {623-637},
year = {2000},
issn = {0168-9002},
doi = {https://doi.org/10.1016/S0168-9002(00)00337-5},
url = {https://www.sciencedirect.com/science/article/pii/S0168900200003375},
author = {R. Dossi and A. Ianni and G. Ranucci and O.Ju. Smirnov}
}

@article{strite,
    author = {Strite, S. and Morko\c{c}, H.},
    title = {GaN, AlN, and InN: A review},
    journal = {Journal of Vacuum Science \& Technology B: Microelectronics and Nanometer Structures Processing, Measurement, and Phenomena},
    volume = {10},
    number = {4},
    pages = {1237-1266},
    year = {1992},
    month = {07},
    issn = {1071-1023},
    doi = {10.1116/1.585897},
    url = {https://doi.org/10.1116/1.585897},
    eprint = {https://pubs.aip.org/avs/jvb/article-pdf/10/4/1237/11845940/1237\_1\_online.pdf},
}

@Article{abid,
AUTHOR = {Abid, Idriss and Mehta, Jash and Cordier, Yvon and Derluyn, Joff and Degroote, Stefan and Miyake, Hideto and Medjdoub, Farid},
TITLE = {AlGaN Channel High Electron Mobility Transistors with Regrown Ohmic Contacts},
JOURNAL = {Electronics},
VOLUME = {10},
YEAR = {2021},
NUMBER = {6},
ARTICLE-NUMBER = {635},
URL = {https://www.mdpi.com/2079-9292/10/6/635},
ISSN = {2079-9292},
DOI = {10.3390/electronics10060635}
}

@article{Baliga,
doi = {10.1088/0268-1242/28/7/074011},
url = {https://doi.org/10.1088/0268-1242/28/7/074011},
year = {2013},
month = {jun},
publisher = {IOP Publishing},
volume = {28},
number = {7},
pages = {074011},
author = {Baliga, B Jayant},
title = {Gallium nitride devices for power electronic applications},
journal = {Semiconductor Science and Technology}
}

@ARTICLE{sequeira,
       author = {{Sequeira}, Miguel C. and {Mattei}, Jean-Gabriel and {Vazquez}, Henrique and {Djurabekova}, Flyura and {Nordlund}, Kai and {Monnet}, Isabelle and {Mota-Santiago}, Pablo and {Kluth}, Patrick and {Grygiel}, Clara and {Zhang}, Shuo and {Alves}, Eduardo and {Lorenz}, Katharina},
        title = "{Unravelling the secrets of the resistance of GaN to strongly ionising radiation}",
      journal = {Communications Physics},
         year = 2021,
        month = dec,
       volume = {4},
       number = {1},
          eid = {51},
        pages = {51},
          doi = {10.1038/s42005-021-00550-2},
       adsurl = {https://ui.adsabs.harvard.edu/abs/2021CmPhy...4...51S}
}

\end{document}


\section*{Supplementary Material}
\subsection{PMT Calibration and SPE Modeling}
\begin{figure}[h!]
    \centering
    \includegraphics[width=0.8\linewidth]{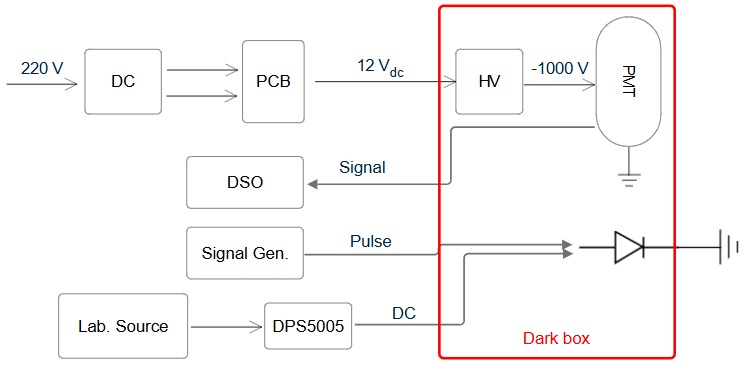}
    \caption{Experimental setup for the PMT calibration. The custom circuit allows for simultaneous modulation of pulse timing and light intensity, synchronized with the DSO for triggered data acquisition.}
    \label{fig:spe}
\end{figure}

The characterization of scintillator light yield requires a precise calibration of the photodetection system to relate the measured charge spectra to the number of detected photons \cite{rossipmt}. This was achieved using a Hamamatsu R6350 PMT calibrated in single photoelectron (SPE) mode. For the calibration setup, a 460 nm LED (L-7113QC-D, Kingbright \cite{led}) was positioned directly in front of the PMT inside a light-tight dark box. The LED was driven by a custom circuit (Figure \cref{fig:spe}) that utilizes two sources: a pulse signal for on/off switching and a direct current (DC) level to modulate light intensity. This configuration, controlled by a DPS5005 \cite{dps5005} power supply, ensured the LED operated at an input voltage of $8.66 V$, maintaining the PMT in the SPE regime at a bias of -1000 V. To improve the signal-to-noise ratio and reduce background triggers, the circuit sent a synchronization pulse to a digital storage (DSO), ensuring data acquisition occurred exclusively during light emission.

\begin{figure}[h!]
    \centering
    \includegraphics[width=0.8\linewidth]{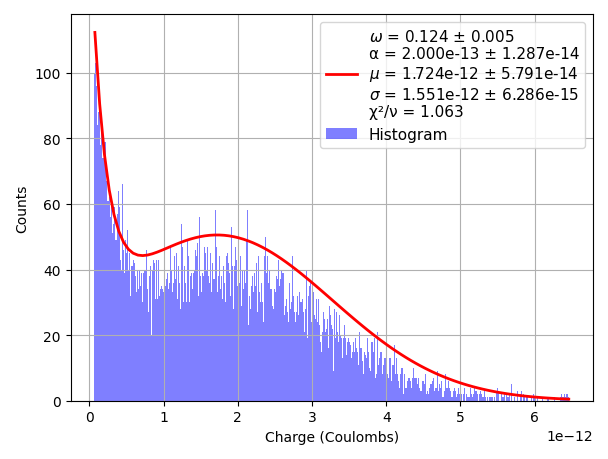}
    \caption{PMT's single photoelectron response function showing the charge spectrum and the model fit.}
    \label{fig:spehsit}
\end{figure}

The resulting charge spectra were modeled using the SPE response function \cref{eq:s0} shown in Figure \cref{fig:spehsit}. This function is defined as:

\begin{equation}
\label{eq:s0}
S(x) = \left( \frac{\omega}{\alpha} e^{-\frac{x}{\alpha}} + \frac{(1-\omega)}{g_N} \frac{1}{\sqrt{2\pi\sigma}} e^{-\frac{(x-\mu)^2}{2\sigma^2}} \right) H(x)
\end{equation}

where the Heaviside step function $H(x)$ ensures $S(x) = 0$ for $x < 0$. The model accounts for two physical processes: first, photoelectrons that backscatter and miss the initial amplification stage are represented by an exponential decay with characteristic width $\alpha$ and event fraction $\omega$. Second, photoelectrons undergoing full amplification through the dynode chain follow a truncated Gaussian distribution with mean $\mu$ and standard deviation $\sigma$. The normalization factor \cref{eq:gn} ensures the Gaussian component integrates to unity:

\begin{equation}
\label{eq:gn}
g_{N} = \frac{1}{2} \operatorname{erfc}\left(-\frac{\mu}{\sqrt{2}\sigma}\right)
\end{equation}

By fitting the measured spectra with \cref{eq:s0}, we determined the per-photon charge at the calibration wavelength (460 nm). Since the light yield depends on the specific emission of each material, this value was subsequently corrected by integrating the PL spectrum of each scintillator weighted by the PMT's quantum efficiency. This procedure yields the effective mean charge per photon received for each specific sample. Finally, for absolute light yield determination, these calibrated responses were compared to the charge spectra obtained by irradiating the samples with a mixed $\alpha$-source ($^{239}$Pu+$^{241}$Am+$^{244}$Cm, 0.07 $\mu$Ci).

\subsection{SRIM}

\begin{figure}[h!]
    \centering
    \includegraphics[width=1\linewidth]{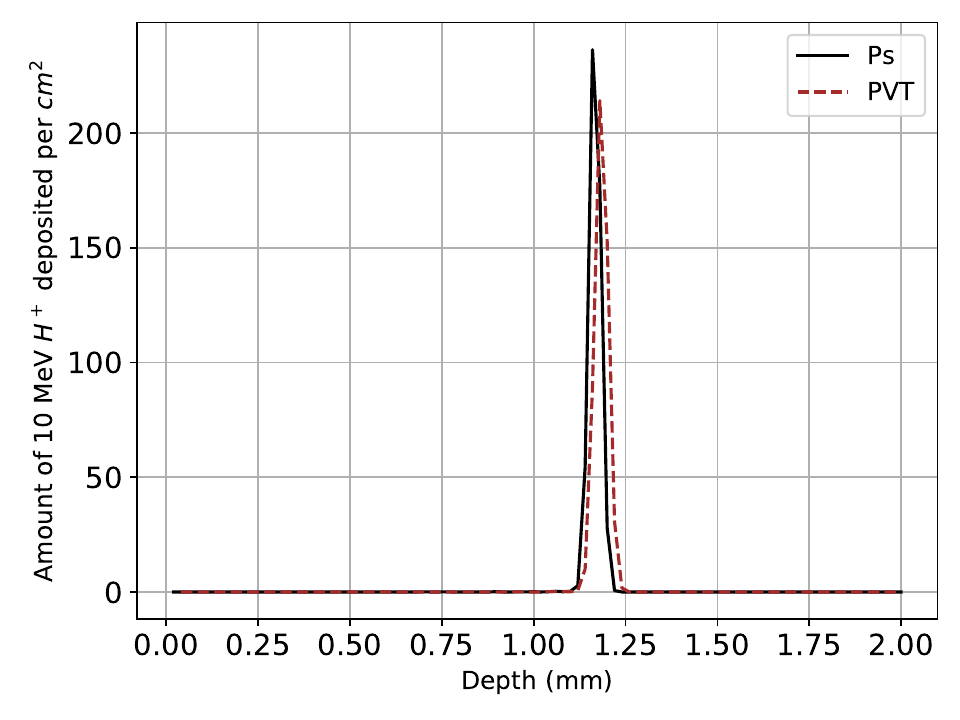}
    \caption{Proton energy deposition in the main plastic scintillator matrices obtained from SRIM simulations. The distributions shown correspond to 10 MeV protons; additional simulations were performed for higher proton energies up to 13 MeV, which is the maximum energy fully contained within the 2 mm scintillator thickness.}
    \label{fig:srim}
\end{figure}

Stopping and Range of Ions in Matter (SRIM) is a widely used software package for simulating ion penetration, energy loss, and range in matter. As shown in \cref{fig:srim}, simulations indicate that protons with an energy of $10\ \mathrm{MeV}$, the energy used in the experiment, are stopped at an average depth of approximately $1.2\ \mathrm{mm}$ within the plastic scintillators. Given the scintillator thickness of about $2\ \mathrm{mm}$, the proton beam is therefore fully absorbed within the material, ensuring that no particles traverse the scintillator.

\clearpage
\bibliography{bib}